\documentclass{article}

\usepackage{arxiv}

\usepackage[utf8]{inputenc} 
\usepackage[T1]{fontenc}    
\usepackage{hyperref}       
\usepackage{url}            
\usepackage{booktabs}       
\usepackage{amsfonts}       
\usepackage{nicefrac}       
\usepackage{microtype}      
\usepackage{lipsum}

\usepackage{graphicx}
\usepackage{dcolumn}
\usepackage{bm}
\usepackage{amsmath}
\graphicspath{{Figs/}}
\usepackage{subfigure}
\usepackage{epstopdf}
\usepackage{csvsimple,longtable,booktabs}
\usepackage{array}
\usepackage{tabularx}
\newcolumntype{C}[1]{>{\centering\let\newline\\\arraybackslash\hspace{0pt}}m{#1}}

\usepackage{hyperref}
\usepackage[mathlines]{lineno}
\usepackage{color,soul}

\usepackage{enumerate}
\usepackage{enumitem}
\usepackage{newlfont}
\usepackage{tikz-network}
\usepackage{tikz}

\newcommand{\fourcycle}[1]{%
	\begin{tikzpicture}[#1]%
	\draw (0,0) -- (0,1ex);%
	\draw (0,0) -- (0.75ex,1ex);%
	\draw (0.75ex,1ex) -- (0.75ex,0);%
	\draw (0.75ex,0) -- (0,1ex);%
	\end{tikzpicture}%
}

\newcommand{\sixcycle}[1]{%
	\begin{tikzpicture}[#1]%
	\draw (0,0) -- (0,1ex);%
	\draw (0,0) -- (0.75ex,1ex);%
	\draw (0.75ex,1ex) -- (1.5ex,0);%
	\draw (1.5ex,0) -- (1.5ex,1ex);%
	\draw (1.5ex,1ex) -- (0.75ex,0);%
	\draw (0.75ex,0) -- (0,1ex);%
	\end{tikzpicture}%
}

\def\genbox#1#2#3#4#5#6{
	\leavevmode\raise#4bp\hbox to#5bp{\vrule height#5bp depth0bp width0bp
		\pdfliteral{q .5 w \csname #2COLOR\endcsname\space RG
			\csname #3PDF\endcsname{#5}{#6} S Q
			\ifx1#1 q \csname #2COLOR\endcsname\space rg 
			\csname #3PDF\endcsname{#5}{#6} f Q\fi}\hss}}

\def\trianbox   #1#2{\genbox{#1}{#2}  {trian}    {0}   {5}    {2.5}}

\title{Transitivity and degree assortativity explained: The bipartite structure of social networks}

\author{
  Demival Vasques Filho \\
  Digital Humanities Lab\\
  Leibniz-Institut f\"ur Europ\"aische Geschichte\\
  Alte Universit\"atsstra{\ss}e 19, 55116 Mainz, Germany\\
  \texttt{vasquesfilho@ieg-mainz.de} \\
   \And
 Dion R.J. O'Neale \\
	Department of Physics, University of Auckland\\
	Te P\={u}naha Matatini, University of Auckland\\ 
	Private Bag 92019, Auckland, New Zealand \\
  \texttt{d.oneale@auckland.ac.nz} \\
}

\begin{document}
\maketitle

\begin{abstract}
Dynamical processes, such as the diffusion of knowledge, opinions, pathogens, ``fake news'', innovation, and others, are highly dependent on the structure of the social network on which they occur. However, questions on why most social networks present some particular structural features, namely high levels of transitivity and degree assortativity, when compared to other types of networks remain open. First, we argue that every one-mode network can be regarded as a projection of a bipartite network, and show that this is the case using two simple examples solved with the generating functions formalism. Second, using synthetic and empirical data, we reveal how the combination of the degree distribution of both sets of nodes of the bipartite network --- together with the presence of cycles of length four and six --- explains the observed levels of transitivity and degree assortativity in the one-mode projected network. Bipartite networks with top node degrees that display a more right-skewed distribution than the bottom nodes result in highly transitive and degree assortative projections, especially if a large number of small cycles are present in the bipartite structure.
\end{abstract}


Real-world networks of distinct nature (e.g. social, biological, technological) usually display common topological features, as heterogeneous degree distribution and small average path length. However, it has been observed that social networks tend to present higher levels of transitivity and degree assortativity than other non-social networks \cite{newman2002assortative,newman2003mixing}. These two structural properties have been extensively addressed in the fields of network theory and social network analysis due to their importance, for instance, in studies investigating spreading of social phenomena. Nonetheless, the reasons that make social networks different are still not fully understood. 

Previous works have conjectured that high transitivity and assortativity are a result of building most social networks using group-based methods \cite{newman2003social,fisher2017perceived} --- which we can directly translate to bipartite networks, with the agents (bottom nodes) belonging to groups (top nodes). Newman and Park \cite{newman2003social} proposed a model and solved it using the generating functions approach to support this idea, while recently, Fisher \textit{et al.} 
\cite{fisher2017perceived} used empirical evidence to do so. The latter gathered information from 88 networks of several types and classified them into three categories: direct (naturally one-mode) social networks, group-based (bipartite) social networks, and others (non-social). They observed that ``\ldots while social networks are more assortative than non-social networks, only when they are built using group-based methods do they tend to be positively assortative.''  

Here, we build on works by Newman and Park \cite{newman2003social}, Estrada \cite{estrada2011combinatorial}, and on a previous work of ours \cite{vasques2018degree} to reveal the effects of the bipartite structure on transitivity and degree assortativity of projections. Transitivity is always present in projected networks --- with one exception, that we will see later --- with levels that depend on the topology of the bipartite network. Concerning assortativity, we provide an answer to the question raised in the conclusions of Larremore \textit{et al.} \cite{larremore2014efficiently}, ``... whether assortativity is due to properties of social networks or due to implicitly projecting from bipartite data...'', showing that degree assortativity is indeed a property of social networks. Moreover, we explain the reasons for the aforementioned observations of Fisher \textit{et al.} \cite{fisher2017perceived} --- that social networks are more degree assortative than other networks --- by arguing that social networks have a inherent group-based structure.


We use empirical networks constructed from the ArXiv and the APS (American Physical Society) datasets, and data collected from Norwegian boards of directors \cite{seierstad2011few}, with a corresponding bipartite configuration model (BiCM) as the null model. Moreover, with the generating functions formalism, we analytically show that every network can be regarded as having an original bipartite structure, even those we refer to as naturally one-mode networks. We will see that the latter is just a projection of a trivial case of bipartite networks. The concept of social networks being originally bipartite is not new. Newman and Park \cite{newman2003social} were the first to propose that social networks inherently have a bipartite structure, followed by Guillaume and Latapy  \cite{guillaume2004bipartite}, who went further to propose that all networks can be modeled as such. 

A bipartite network is a graph $B = \{U,V,E\}$, where $U$ and $V$ are disjoint sets of nodes and $E = \{(u,v):u \in U, v \in V\}$ is the set of links connecting nodes of the different sets. We refer to the sets $U$ and $V$ as the bottom and top node sets, respectively. A projection onto the nodes $U$ (a so-called bottom projection) results in a one-mode network, where node $u$ is connected to $u'$, $\{{u,u'}\}$ $\in$ $U$, only if there exists at least one common neighbor $v$ $\in$ $V$, that is, at least a pair of edges $(u,v)$ and $(u',v)$ $\in$ $E$. The converse is also valid:  every analysis for projected networks onto the bottom set of nodes $U$ works also for top projections onto $V$. Hence, for the sake of simplicity, we treat all projections as bottom projections. That is, we define a projected network as a graph $G = \{U,L\}$, where $L$ is the set of links in the projection, such that $L = \{(u,u'):\exists\, v \in V, (u,v) \in E\, \textrm{and}\, (u',v) \in E\}$. Projections have different flavors. Here, we mention simple graph $G_{\textrm{s}}$, (simple) weighted graph $G_{\textrm{w}}$, and multigraph $G_{\textrm{m}}$ projections \cite{vasques2018degree}. Then, we have, for a  bipartite network

\begin{center}
	$k_{u}$ --- degree of bottom node $u$ $\in$ $U$;\\
	$d_{v}$ --- degree of top node $v$ $\in$ $V$;\\
	$P_{\textrm{b}}(k)$ --- bottom node degree distribution;\\
	$P_{\textrm{t}}(d)$ --- top node degree distribution.
	\vspace{0.2cm}
\end{center}  

And, for the projected network,

\begin{center}
	$q_u$ --- degree of node $u \in U$;\\ 
	$P(q)$ --- projected node degree distribution.\\
	\vspace{0.2cm}
\end{center} 

The relation between node degrees in $B$ and $G$ is given by \cite{vasques2018degree}
\begin{equation}
\label{eq:qi}
q_{u} \leq \sum_{j=1}^{k_{u}}(d_{v_j} - 1)\,,
\end{equation}
where the equality holds for $q^{\textrm{m}}_{u}$, the degree of $u$,  when the projection is a multigraph projection. The multigraph degree is also known as  node strength \cite{barrat2004architecture}, $s_{u}$, such that $q^{\textrm{m}}_{u} = s_{u} = \sum_{u'} w_{uu'}$, where $w_{uu'}$ is the weight of the link connecting nodes $u$ and $u'$, in a simple weighted projection (for other weighted projection methods, see \cite{coscia2019impact}). 

In \cite{vasques2018degree} we have shown that using the generating functions formalism, we can predict the node strength (multigraph degree) distribution of a projected network, by knowing the degree distribution of both sets of nodes of the bipartite network. To illustrate our approach in that paper, we solved two very simple cases of degree distributions for bipartite networks, that we recall here. For both examples, we assume that the degree distribution of top nodes, $P_{\textrm{t}}(d)$, follows a delta function distribution, that is, every top node has the same degree $d_{v} = d^{*}$. In the first case, the bottom degree distribution, $P_{\textrm{b}}(k)$, follows a Poisson distribution. Then, according to Eq. (31) of \cite{vasques2018degree}, the node strength distribution $P(q^{\textrm{m}})$ of the projected network is calculated according to

\begin{equation}
\ P(q^{\textrm{m}}) = 
\begin{cases}
\frac{\langle k \rangle ^ {\left(\frac{q^{\textrm{m}}}{d^{*}-1}\right)} e^{-\langle k \rangle}}{\left(\frac{q^{\textrm{m}}}{d^{*}-1}\right)!} & \text{if $(d^{*}-1)|q^{\textrm{m}}$}\\
0 & \text{otherwise}\,.
\end{cases} 
\label{eq:deltaPoisson}
\end{equation}

For the second case, the bottom node degrees are exponentially distributed. The projected strength distribution is 

\begin{equation}
\ P(q^{\textrm{m}}) = 
\begin{cases}
\lambda e^{-\lambda \frac{q^{\textrm{m}}}{d^{*}-1}} & \text{if $(d^{*}-1)|q^{\textrm{m}}$}\\
0 & \text{otherwise}\,,
\end{cases} 
\label{eq:deltaexp}
\end{equation}

which is the Equation (34) of \cite{vasques2018degree}, and where $\lambda = \langle k \rangle^{-1}$.

We can see from Eqs. (\ref{eq:deltaPoisson}) and (\ref{eq:deltaexp}) that the expected multigraph degree distribution for the projected networks also follow Poisson and exponential distributions, respectively (as the bottom nodes), modulated by $d^{*}-1$. If we take $d^{*}=2$, then we have the exact bottom degree distribution. In other words, direct (naturally one-mode) networks are projections of a trivial bipartite network, in which all top nodes of $B$ have degree $d^{*}=2$. We can deem the top nodes as events that create links in a network as, for instance, the exchange of a letter between two persons, the acceptance of a Facebook friendship request, or a trade agreement between two countries.

In the above examples, we have chosen Poisson and exponential distributions as bottom degree distributions due to their simple analytical solutions via generating functions. However, this reasoning can be extended to any other distribution, as well as to simple and weighted projections. In these cases, the resulting degree distribution of the projected network depends also on a feature frequently observed in real-world bipartite networks: four-cycles, that is, pairs of bottom nodes that share more than one (top node) common neighbor.  

Four-cycles are a motif in $B$ that affect the degree of nodes in simple and weighted projections \cite{vasques2019bipartite}. The degree of $u$ is the same in these two types of projection, (i.e. $q^{\textrm{s}}_{u} = q^{\textrm{w}}_{u}$), and represents the number of neighbors $|N(u)|$ of $u$ in $G$. The inequality of Eq. \ref{eq:qi} depends on the number of these cycles of which $u$ is part and on which nodes form the cycles together with $u$. In other words, it is not possible to solve the inequality of Eq. \ref{eq:qi} just by counting the number of four-cycles in the bipartite network. This type of motif is directly related to the concept of redundancy \cite{latapy2008basic}. Four-cycles generate redundant links when creating a simple graph projection (or adds to the weight of the link, when the projection is weighted).

From now on, we will treat all projections as a simple weighted projection, unless otherwise stated. We will use $q^{\textrm{s}}_{u}$ as the node degree and $q^{\textrm{m}}_{u}$ as the node strength, as a reminder of the simple graph degree and the multigraph degree, respectively.

Another motif relevant for the topology of bipartite networks, and the resulting topology of projections, is the cycle of length six, that we refer to as six-cycle. While four-cycles create redundancy of links, six-cycles result in triadic closure in the projected network \cite{opsahl2013triadic,vasques2019bipartite}. That is, one-mode projections of bipartite networks have two types of triangles: those induced by top nodes with degree $d \geq 3$, and those induced by six-cycles \cite{opsahl2013triadic,vasques2019bipartite}. Then, the number of triangles in a projected network is given by
\begin{equation}
\label{eq:triangles}
|\trianbox0{cblack}| \leq \sum_{\substack{v=1 \\ d_{v}>2}}^{|V|} \binom{d_{v}}{3} + |\sixcycle{scale=1.4}| \,,
\end{equation}
where $\sixcycle{scale=1.4}$ is the set of six-cycles in the bipartite network.

Let us recall that the level of transitivity (global clustering coefficient) in networks is calculated as the ratio of the number of triangles to the number of open triplets in the network, according to
\begin{equation}
\label{eq:global_clustering}
C = \dfrac{\text{\# closed triplets}}{\text{\# open triplets}} = \frac{3\times |\trianbox0{cblack}|}{|\wedge|}\,,
\end{equation} 
where $\trianbox0{cblack}$ is the set of triangles and $\wedge$ is the set of open triplets in the network. The higher the presence of six-cycles in the bipartite network, the more triangles can be formed in the projection, increasing transitivity. However, quantifying this relationship is not a simple task and depends on a combinations of factors, similarly to the relationship between four-cycles and the projected degree distribution.

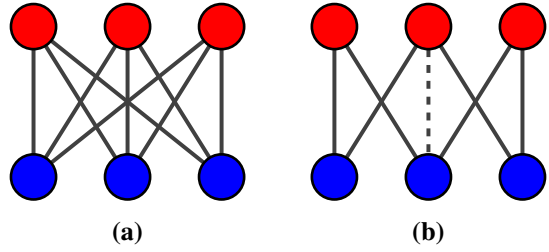
\begin{figure}
	\centering
	
	\begin{tikzpicture}
	
	\Vertex[color=red,position=above]{p1}
	\Vertex[x = 1.25,color=red]{p2}
	\Vertex[x = 2.5,color=red]{p3}
	
	\Vertex[y=-2,color=blue]{a1}
	\Vertex[y=-2,x = 1.25,color=blue]{a2}
	\Vertex[y=-2,x = 2.5,color=blue]{a3}
	
	\Edge(p1)(a1)
	\Edge(p1)(a2)
	\Edge(p1)(a3)
	
	\Edge(p2)(a1)
	\Edge(p2)(a2)
	\Edge(p2)(a3)
	
	\Edge(p3)(a2)
	\Edge(p3)(a1)
	\Edge(p3)(a3)
	
	\Text[x=1.25,y=-2.75]{\textbf{(a)}}

	\Vertex[x = 4,color=red,position=above]{c1}
	\Vertex[x = 5.25,color=red]{c2}
	\Vertex[x = 6.5,color=red]{c3}
	
	\Vertex[y=-2,x = 4,color=blue]{b1}
	\Vertex[y=-2,x = 5.25,color=blue]{b2}
	\Vertex[y=-2,x = 6.5,color=blue]{b3}
	
	\Edge(c1)(b1)
	\Edge(c1)(b2)
	
	\Edge(c2)(b1)
	\Edge[style={dashed}](c2)(b2)
	\Edge(c2)(b3)
	
	\Edge(c3)(b2)
	\Edge(c3)(b3)
	
	
	\Text[x=5.25,y=-2.75]{\textbf{(b)}}
	
	\end{tikzpicture}
	
	\caption{Diagram with example of structures that inhibit the formation of (simple) triangles. The six-cycles are embedded in subgraphs containing top nodes with degree $d \geq 3$. In (a) we have a complete subgraph with an overlapping of nodes with $d_{v} = 3$ that results in only one triadic closure in the projected network, as for instance, three papers co-authored by the same three authors. In (b) the addition of the dashed link breaks the six-cycle into four-cycles.}
	\label{fig:inhibt}
\end{figure}

The inequality of Eq. \ref{eq:triangles} is due to several patterns that emerge in real-world bipartite networks as, for instance, complete subgraphs and the overlapping of four cycles with six-cycles (Fig. \ref{fig:inhibt}). Six-cycles are embedded in such patterns, resulting in a smaller number of triangles in simple projection clustering calculations (i.e those without weighted or multi links) than we would expect. 

Yet, Eq. \ref{eq:triangles} is the answer to why most social networks have relatively high levels of transitivity. The number of triangles is highly dependent on  the node degree distribution $P_{\textrm{t}}(d)$. The broader the top degree distribution, the higher the number of triangles, and therefore the larger the values of transitivity. However, the level of transitivity in a network also depends on the number of open triplets (Eq. \ref*{eq:global_clustering}). The latter, in turn, is related to $P_{\textrm{b}}(k)$. The broader the bottom degree distribution, the more open triplets can be found in the projection. Table \ref{tb:trans_synthetic} shows the number of triangles and open triplets in projections of several synthetic bipartite configuration model (BiCM) networks. While top nodes create cliques in the projected network, bottom nodes connect these cliques together. Then, networks that are naturally one-mode, if modeled as projections of bipartite networks (i.e. $P_{\textrm{t}}(d)$ is a delta distribution with $d^{*}=2$), would have triangles created exclusively by six-cycles. In a case where $d^{*}=2$ and $|\sixcycle{scale=1.4}| \rightarrow 0$, we have the special case of bipartite networks that result in one-mode projections with $C \rightarrow 0$.  

\begin{table*}
	\vspace{0.1cm}
	\centering
	\caption{Mean (standard deviation) values of properties for BiCM networks and their projections, over 100 runs. These networks are created with the same methods as in \cite{vasques2018degree}, where the degree sequences are drawn from probability distributions.  Here, we use delta, Poisson and exponential distribution for both bottom and top nodes. We use $|U| \approx |V| \approx 1,000$, and $\langle k \rangle = \langle d \rangle = 5$. The upper bound number of triangles is the sum of the first and second columns, according to Eq. \ref{eq:triangles}. $|\trianbox0{cblack}|$ is the actual number of triangles, $|\wedge|$ is the number of open triplets, and $C$ is the transitivity of the projection. Note that as $P_{t}(d)$ becomes broader, with the same $P_{b}(k)$, the increase in the number of triangles prevails over the increase in the number of open triplets, and the transitivity consequently increases. Otherwise, as $P_{b}(k)$ becomes broader, keeping $P_{t}(d)$ the same, the increase of open triplets prevails, and the transitivity decreases.
		\label{tb:trans_synthetic}}
	\vspace{0.2cm}
	\begin{tabular}{llllll}
		\cline{2-6}
		\multicolumn{1}{l|}{}         & \multicolumn{5}{c|}{$P_{t}(d)$ = Delta}                                                                                                                                                                                                                                 \\ \hline
		\multicolumn{1}{|l|}{$P_{b}(k)$}   & \multicolumn{1}{c|}{$\sum_{v} \binom{d_{v}}{3}$}                & \multicolumn{1}{l|}{$|\sixcycle{scale=1.4}|$}                & \multicolumn{1}{l|}{$|\trianbox0{cblack}|$}             & \multicolumn{1}{l|}{$|\wedge|$}                                                              & \multicolumn{1}{l|}{$C$}        \\ \hline
		\multicolumn{1}{|l|}{Delta}   & \multicolumn{1}{l|}{9,956 (15)}    & \multicolumn{1}{l|}{673 (26)}        & \multicolumn{1}{l|}{10,622 (29)}     & \multicolumn{1}{l|}{186,546 (448)}                                               & \multicolumn{1}{l|}{0.171 (0.001)}   \\
		\multicolumn{1}{|l|}{Poisson} & \multicolumn{1}{l|}{9,834 (97)}   & \multicolumn{1}{l|}{1,298 (68)}       & \multicolumn{1}{l|}{11,094 (146)}     & \multicolumn{1}{l|}{220,181 (4,628)}                                               & \multicolumn{1}{l|}{0.151 (0.001)} \\
		\multicolumn{1}{|l|}{Exponential}    & \multicolumn{1}{l|}{9,750 (166)}  & \multicolumn{1}{l|}{5,114 (717)}       & \multicolumn{1}{l|}{14,105 (628)}    & \multicolumn{1}{l|}{313,217 (14,386)}                                             & \multicolumn{1}{l|}{0.135 (0.001)} \\ \hline
		&                                         &                                               &                                            &                                                                                        &                                    \\ \cline{2-6} 
		\multicolumn{1}{l|}{}         & \multicolumn{5}{c|}{$P_{t}(d)$ = Poisson}                                                                                                                                                                                                                               \\ \hline
		\multicolumn{1}{|l|}{Delta}   & \multicolumn{1}{l|}{20,316 (955)}   & \multicolumn{1}{l|}{1,287 (75)}     & \multicolumn{1}{l|}{21,587 (1,017)}    & \multicolumn{1}{l|}{299,836  (12,276)} & \multicolumn{1}{l|}{0.216 (0.002)} \\
		\multicolumn{1}{|l|}{Poisson} & \multicolumn{1}{l|}{20,517 (1,069)}  & \multicolumn{1}{l|}{2,492 (260)}    & \multicolumn{1}{l|}{22,917 (1,305)}   & \multicolumn{1}{l|}{357,399 (22,096)}                                             & \multicolumn{1}{l|}{0.192 (0.002)} \\
		\multicolumn{1}{|l|}{Exponential}    & \multicolumn{1}{l|}{20,100 (936)} & \multicolumn{1}{l|}{9,705 (1,431)}    & \multicolumn{1}{l|}{28,080 (1,657)}    & \multicolumn{1}{l|}{488,607 (28,643)}                                             & \multicolumn{1}{l|}{0.173 (0.002)} \\ \hline
		&                                         &                                               &                                            &                                                                                        &                                    \\ \cline{2-6} 
		\multicolumn{1}{l|}{}         & \multicolumn{5}{c|}{$P_{t}(d)$ = Exponential}                                                                                                                                                                                                                           \\ \hline
		\multicolumn{1}{|l|}{Delta}   & \multicolumn{1}{l|}{76,420 (10,542)}  & \multicolumn{1}{l|}{5,105 (709)}    & \multicolumn{1}{l|}{81,406 (11,193)}   & \multicolumn{1}{l|}{818,619   (91,143)} & \multicolumn{1}{l|}{0.299 (0.011)}  \\
		\multicolumn{1}{|l|}{Poisson} & \multicolumn{1}{l|}{74,837 (8,724)}  & \multicolumn{1}{l|}{9,556 (1,349)}   & \multicolumn{1}{l|}{83,836 (9,803)}  & \multicolumn{1}{l|}{929,043 (93,213)}                                            & \multicolumn{1}{l|}{0.27 (0.01)}   \\
		\multicolumn{1}{|l|}{Exponential}    & \multicolumn{1}{l|}{77,457 (11,075)} & \multicolumn{1}{l|}{37,490 (11,084)} & \multicolumn{1}{l|}{105,281 (17,670)} & \multicolumn{1}{l|}{1,263,366 (183,203)}                                           & \multicolumn{1}{l|}{0.249 (0.008)} \\ \hline
		\vspace{0.1cm}
	\end{tabular}
\end{table*}

In \cite{estrada2011combinatorial}, Estrada has shown that ``The assortativity of a network depends on the balance between three structural factors: transitivity (clustering), intermodular connectivity, and relative branching.'' The first two contribute positively to the increase of assortativity, while the third contributes negatively. The degree assortativity $r$, in terms of such structural features (Equation (6) of \cite{estrada2011combinatorial}), is given by

\begin{equation}
\label{eq:estrada_r}
r = \frac{|P_{2}|(|P_{3/2}| + C - |P_{2/1}|)}{3|S_{1,3}| - |P_{2}|(|P_{2/1}| - 1)} \,,
\end{equation}
where $S_{1,3}$ is a star subgraph of four nodes, $|P_{2}|$ is the number of paths of size two; $C$ is the transitivity; $|P_{3/2}|$ is the ratio of the number of paths of size three by the number of paths of size two ---  the so-called intermodular connectivity; and $|P_{2/1}|$ is the relative branching. For a more fundamental explanation of these concepts, see \cite{estrada2011combinatorial}.

If the degree distributions and six-cycles of $B$ determine transitivity, they certainly play a part in the projected assortativity. Similarly, if four-cycles affect the projected degree distribution, it is reasonable to assume that they might affect the degree assortativity as well. In \cite{vasques2018degree}, we have already shown that BiCM can generate degree assortative, neutral and dissortative projected networks, different to what was previously believed \cite{ramasco2004self}, depending on the combination of $P_{t}(d)$ and $P_{b}(k)$. However, the BiCM is not enough to fully understand degree assortativity, precisely due to the absence of large numbers of four and six-cycles in random networks when compared to real-world networks \cite{vasques2019bipartite}. Thus, let us now explore the topology of several empirical bipartite networks (i.e bipartite assortativity, degree distributions, and four- and six-cycles) and its relation with transitivity and degree (and strength) assortativity of their projections (i.e. triangles, open triplets, relative branching and intermodular connectivity).

\begin{figure*}
	\centering
	\subfigure{\label{fig:bio} \includegraphics[scale=0.19]{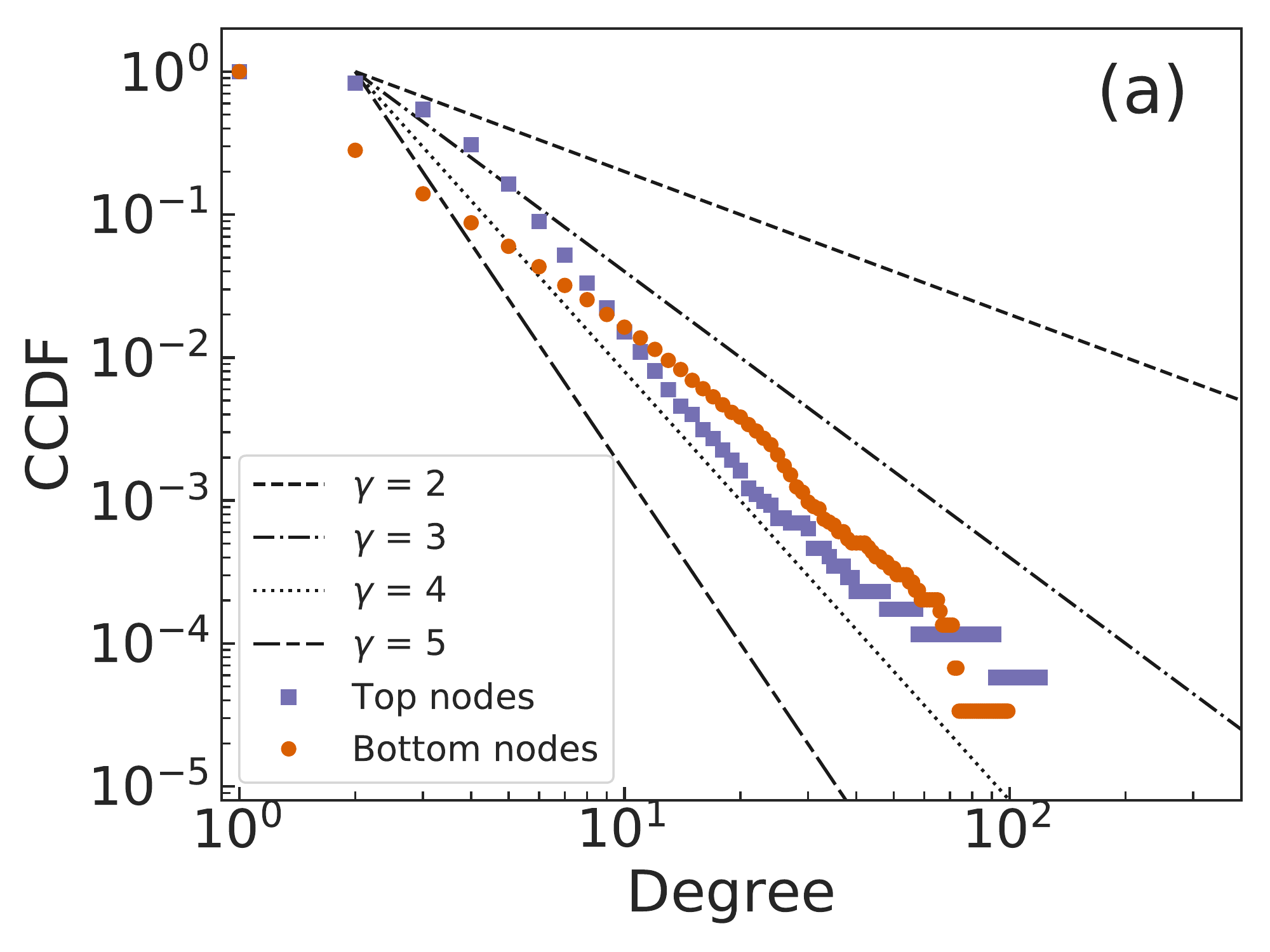}}
	\subfigure{\label{fig:maths} \includegraphics[scale=0.19]{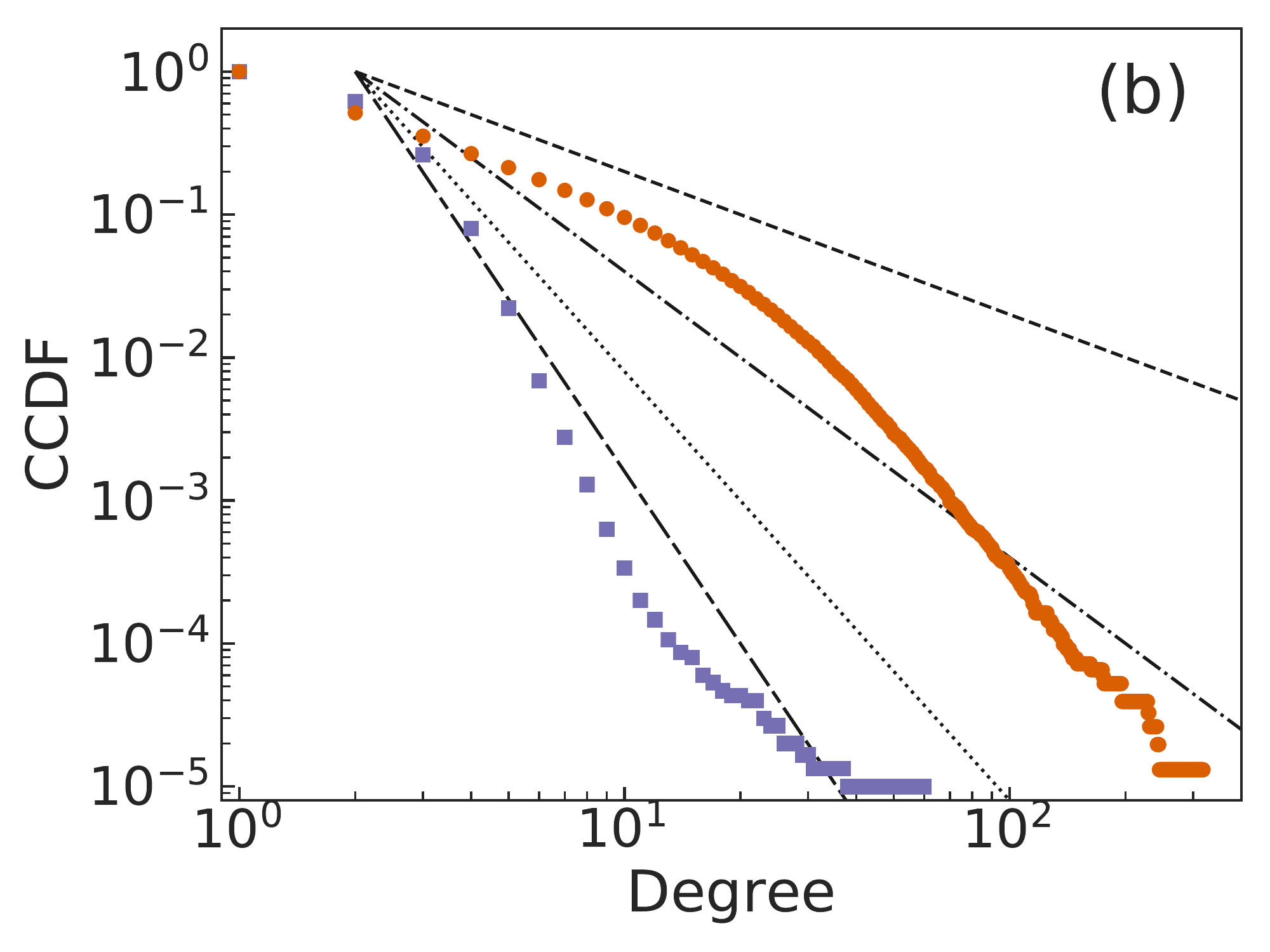}}
	\subfigure{\label{fig:pre} \includegraphics[scale=0.19]{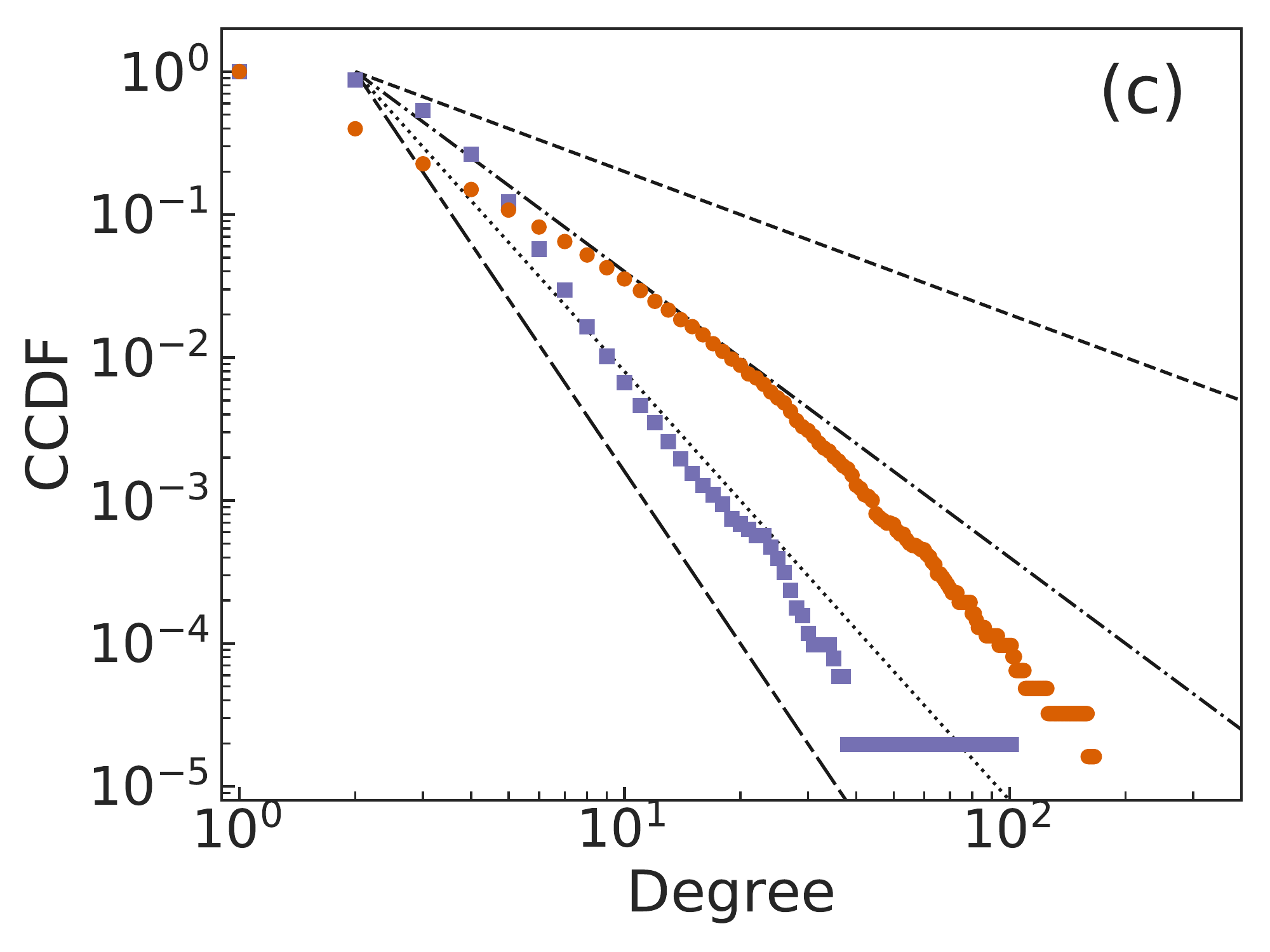}}
	\subfigure{\label{fig:norwegian} \includegraphics[scale=0.19]{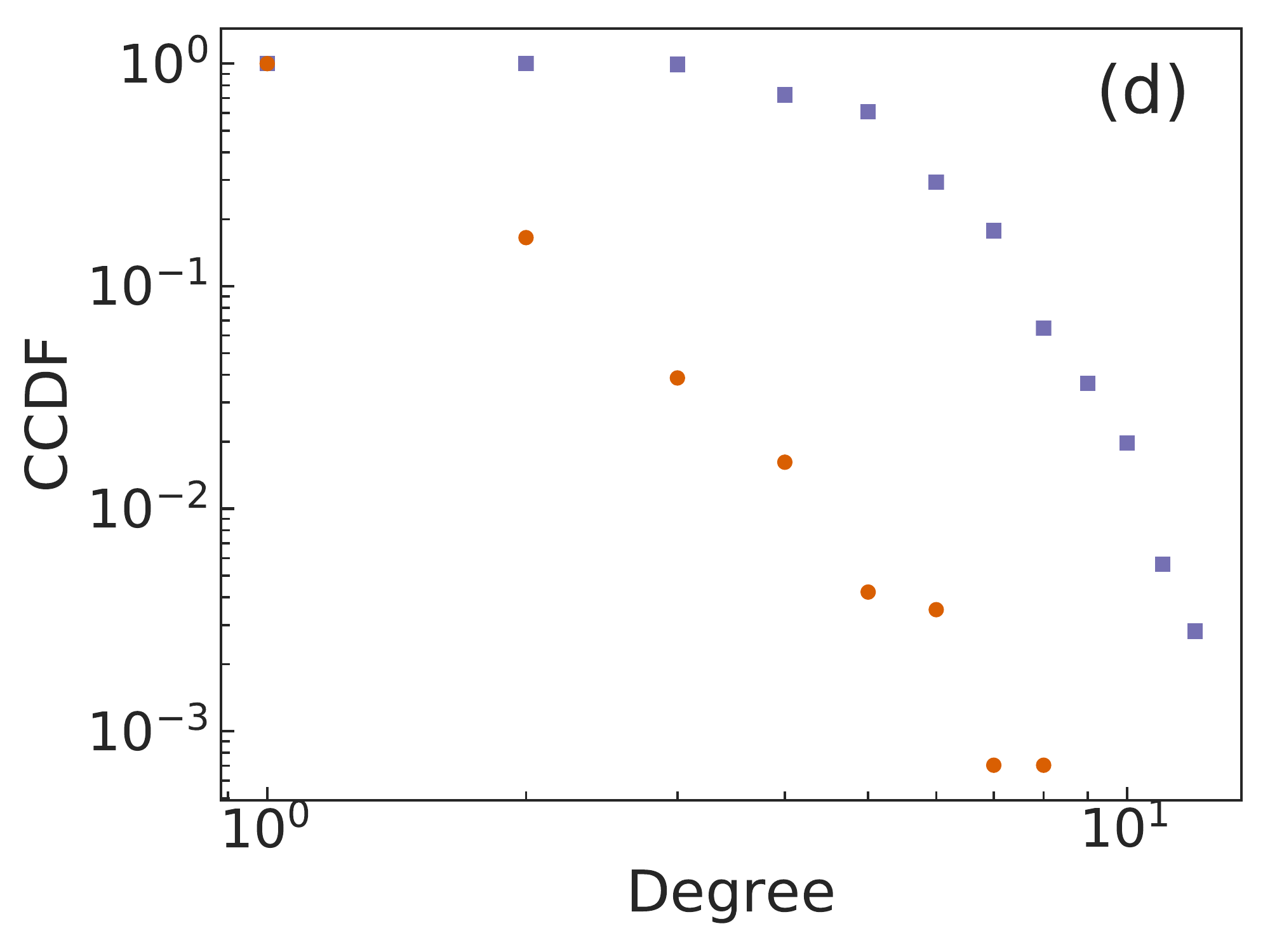}}
	\caption{Degree distributions of top and bottom node sets for: (a) ArXiv Biology, (b) ArXiv Mathematics (c) PRE --- Physical Review E, and (d) Norwegian directors. The heavy-tailed top degree distributions of ArXiv Bio and PRE, supported by the high presence of small cycles (Table \ref{tb:summary}), result in high projected transitivity and degree assortativity. This is so even with both cases displaying heavy-tailed bottom distributions, that create large numbers of open triplets and high relative branching. On the other hand, peaked distributions of the Norwegian board-directors network result in comparatively small projected degree assortativity, but high transitivity as the top distribution is more shifted to the right than the bottom distribution. In this case, transitivity is driving assortativity.}
	\label{fig:degree_distributions}
\end{figure*}

For that, we create three large scientific networks --- connecting authors to papers, for every paper published in ArXiv Biology, ArXiv Mathematics, and Phys. Rev. E (PRE), since their beginning in the 1990s until 2015 --- and a board-affiliation network, which is a snapshot of directors sitting in the board of public limited liability companies in Norway, in 2011 \cite{seierstad2011few}. The bipartite degree distribution of these networks are shown in Fig. \ref{fig:degree_distributions}. We also create their corresponding BiCM, keeping the same degree sequence as the original networks, but rewiring the links at random, following the method we use in \cite{vasquesfilho2019structure}. The structural properties of our interest, of both empirical and BiCM networks, are listed in Table \ref{tb:summary}. 

\begin{table*}
	\vspace{0.2cm}
	\small
	\centering
	\caption{Summary of several structural properties related to transitivity and degree assortativity for the empirical (Emp) networks and their respective configuration model (CM) --- mean (standard deviation) over 100 runs. The BiCM breaks the particular patterns found in empirical social networks. The result is a much larger number of open triplets, and a higher level of relative branching, decreasing both projected transitivity and degree assortativity, respectively, when compared to the empirical projections. As the number of four and six cycles is relatively low in the Norwegian Board network, the differences between the projected empirical and configuration model networks are much more subtle than for the other cases.  \label{tb:summary}}
	\vspace{0.2cm}
	\begin{tabular}{|p{1.3cm}|p{1.2cm}p{1.6cm}|p{1.2cm}p{1.6cm}|p{1.2cm}p{1.6cm}|p{1.2cm}p{1.6cm}|}
		\hline
		& \multicolumn{2}{c|}{ArXiv Biology} & \multicolumn{2}{c|}{ArXiv Mathematics} & \multicolumn{2}{c|}{Physical Review E} & \multicolumn{2}{c|}{Norwegian Board} \\ 
		Properties                 & Emp  & CM        & Emp  & CM         & Emp & CM        & Emp  & CM        \\
		\hline
		$r_{B}$                    & -0.09      & 0.00 (0.00)        & -0.10      & -0.09 (0.00)        & -0.07     & -0.05 (0.00)       & -0.65  & -0.63 (0.01)   \\
		$|\fourcycle{scale=1.4}|$  & 34,659     & 85 (10)            & 515,144    & 148 (13)            & 131,544    & 150 (13)           & 92    & 1.5 (1.2)         \\
		$|\sixcycle{scale=1.4}|$   & 5,272      & 943 (91)           & 263,487    & 2,418 (75)          & 47,877    & 1,895 (99)         & 7      & 2.4 (1.7)        \\
		$r_{G_{\textrm{s}}}$       & 0.88       & 0.37 (0.02)        & 0.31       & 0.03 (0.00)         & 0.50      & 0.10 (0.01)        & 0.22   & 0.14 (0.02)    \\
		$r_{G_{\textrm{m}}}$       & 0.73       & 0.36 (0.02)        & 0.18       & 0.03 (0.00)         & 0.314     & 0.10 (0.01)        & 0.23   & 0.14 (0.02)    \\
		$|\trianbox0{cblack}|$     & 639,571    & 645,541 (4,746)    & 409,855    & 443,998 (965)       & 459,726   & 499,535 (2,687)    & 5,606  & 5,634 (17)     \\
		$|\wedge|$                       & 2,289,465  & 3,915,117 (63,604) & 3,516,171  & 12,865,053 (64,747) & 2,691,848 & 6,516,593 (62,436) & 24,590 & 26,484 (394)   \\
		$C$                        & 0.84       & 0.49 (0.01)        & 0.35       & 0.10 (0.00)         & 0.51      & 0.23 (0.00)        & 0.68   & 0.64 (0.01)    \\
		$|P_{2/1}|$                & 24.5       & 35.9 (0.6)         & 11.5       & 27.2 (0.1)          & 16.2      & 29.2 (0.3)         & 6.4    & 6.7 (0.1)    \\
		$|P_{3/2}|$                & 72.4       & 62.7 (1.3)         & 20.4       & 29.8 (0.4)          & 34.5      & 36.6 (0.8)         & 6.4    & 6.6 (0.2)        \\
		\hline    
	\end{tabular}
\end{table*}

In these four networks, the bipartite assortativity $r_{B}$ is negative and unrelated to the projected degree assortativity $r_{G_{\textrm{s}}}$ and the node strength assortativity $r_{G_{\textrm{m}}}$, which are both positive. Assortativity values are calculated as the Pearson coefficient of the degree-degree correlation \cite{newman2002assortative}. In addition, $r_{G_{\textrm{s}}}$ is also calculated using Eq. \ref{eq:estrada_r}, which gives the same results, as expected. 

\begin{figure*}
	\centering
	\subfigure{\label{fig:bio_bp} \includegraphics[scale=0.19]{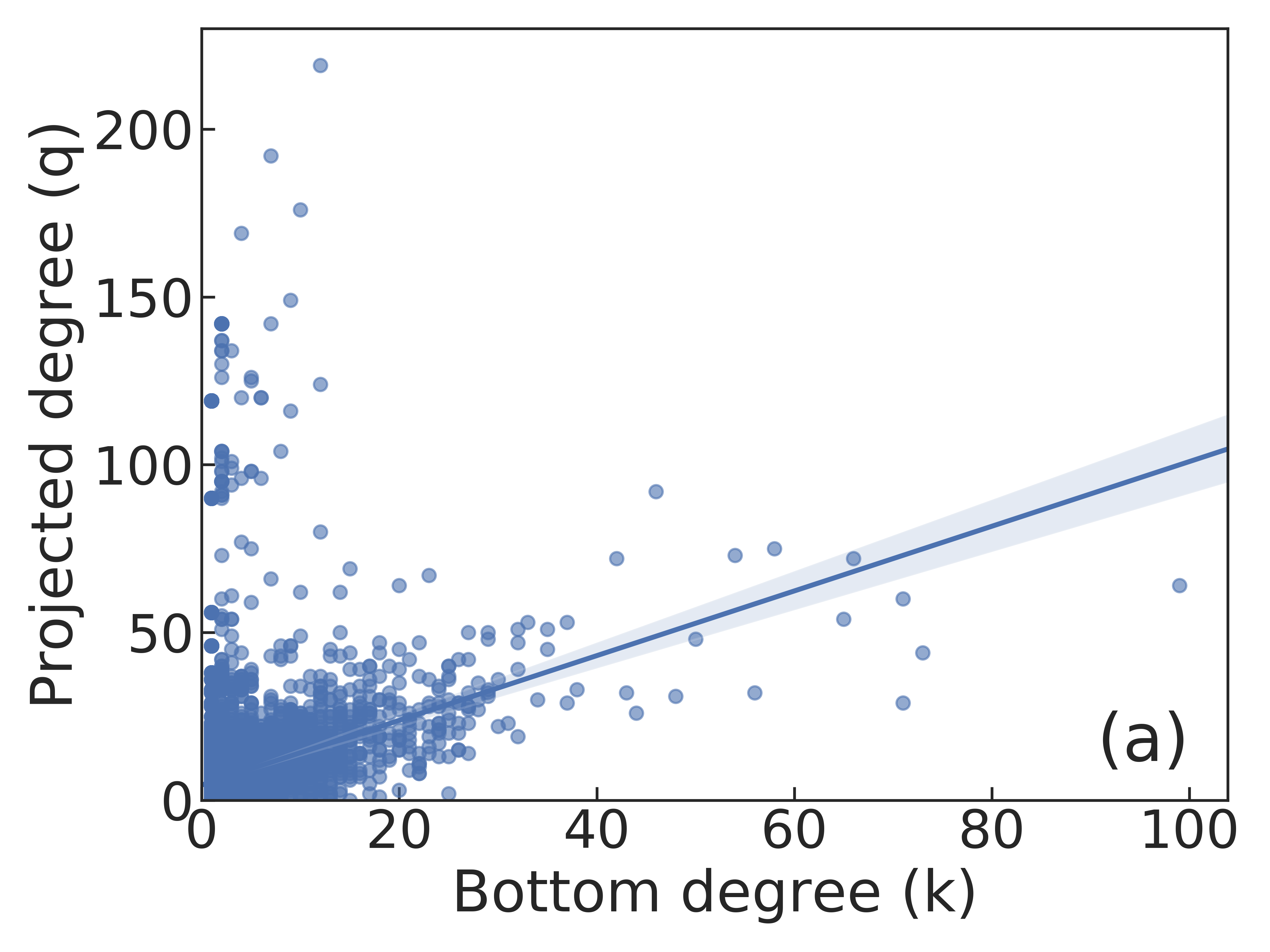}}
	\subfigure{\label{fig:maths_bp} \includegraphics[scale=0.19]{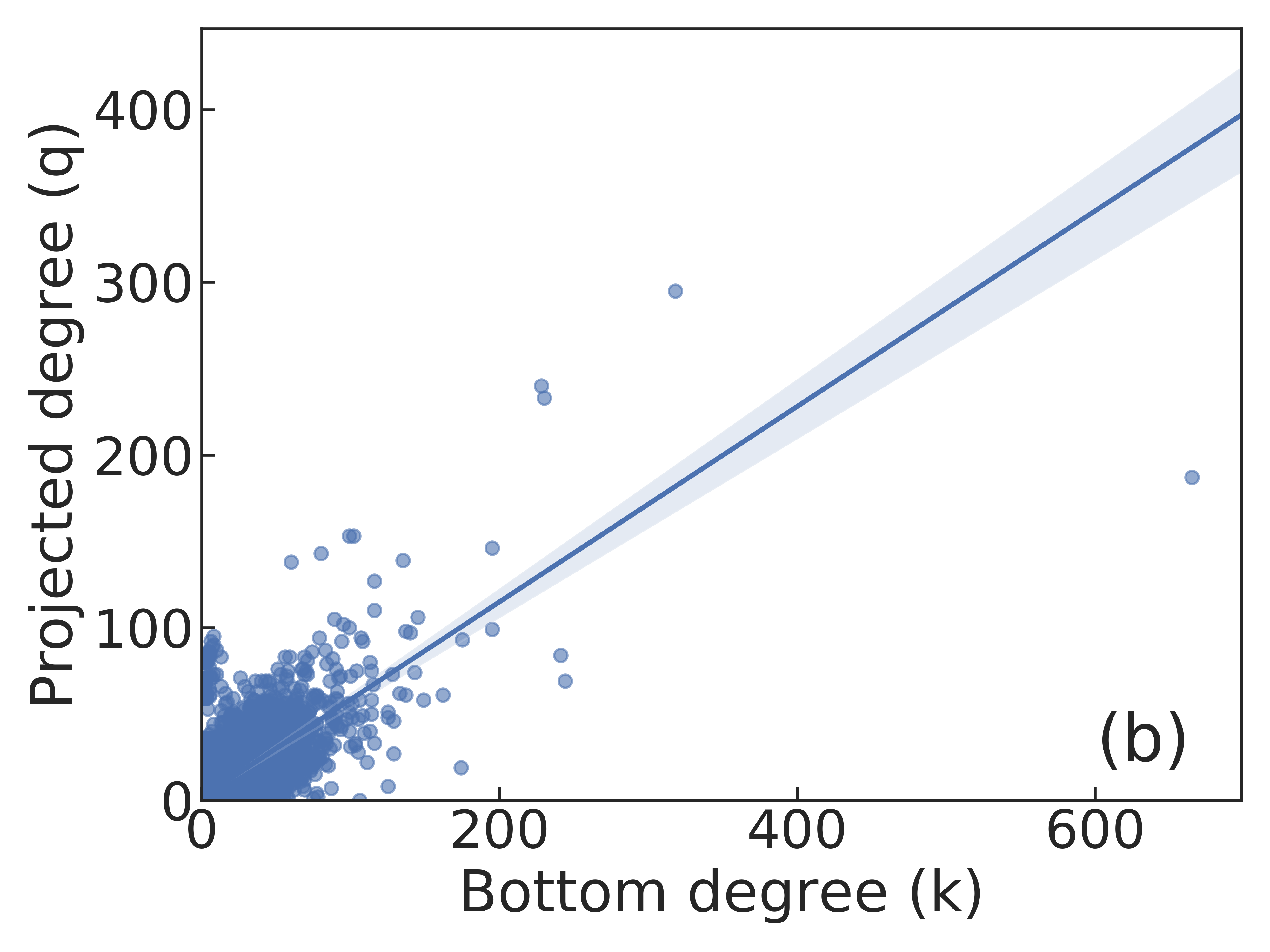}}
	\subfigure{\label{fig:pre_bp} \includegraphics[scale=0.19]{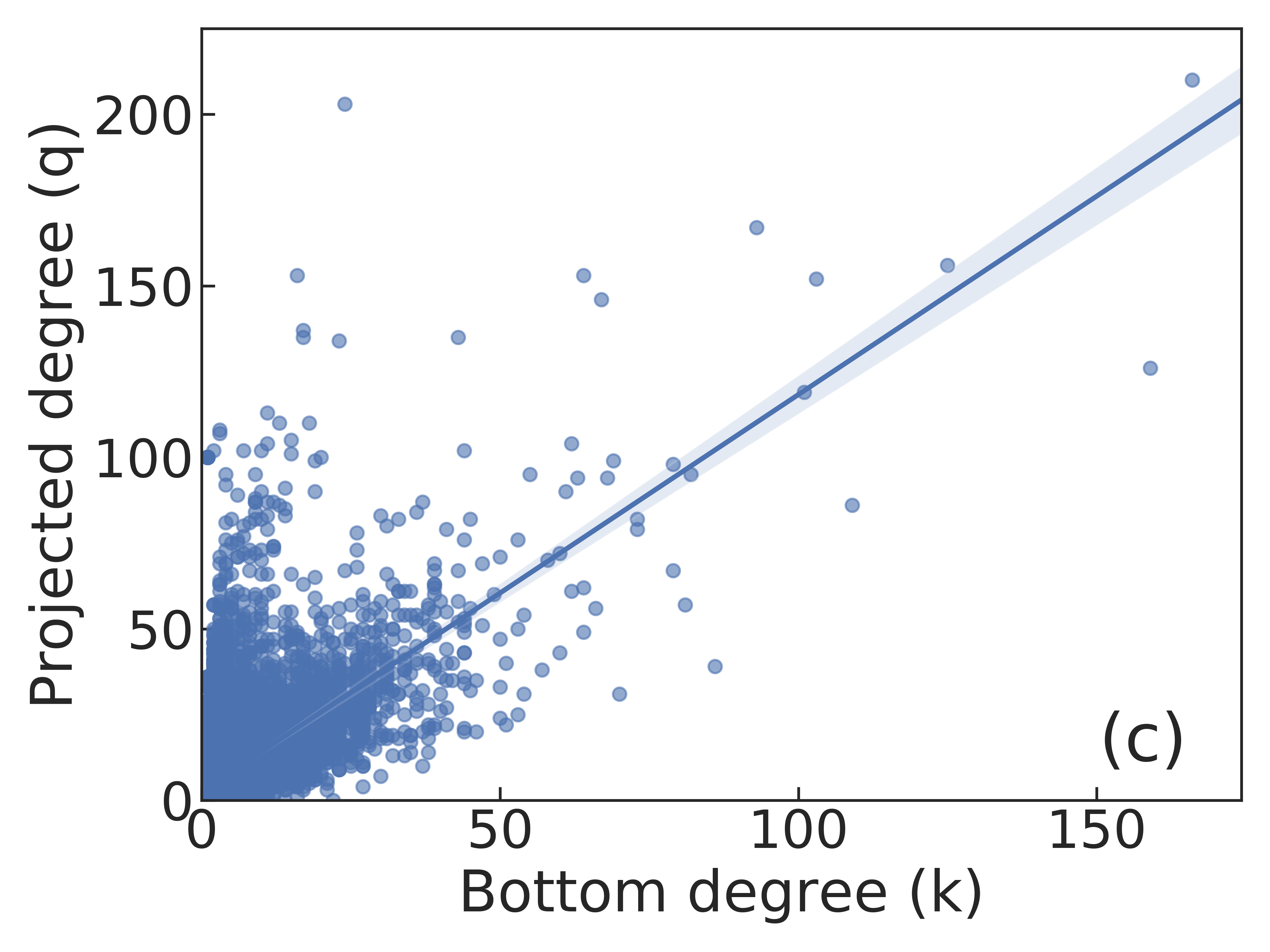}}
	\subfigure{\label{fig:norwegian_bp} \includegraphics[scale=0.19]{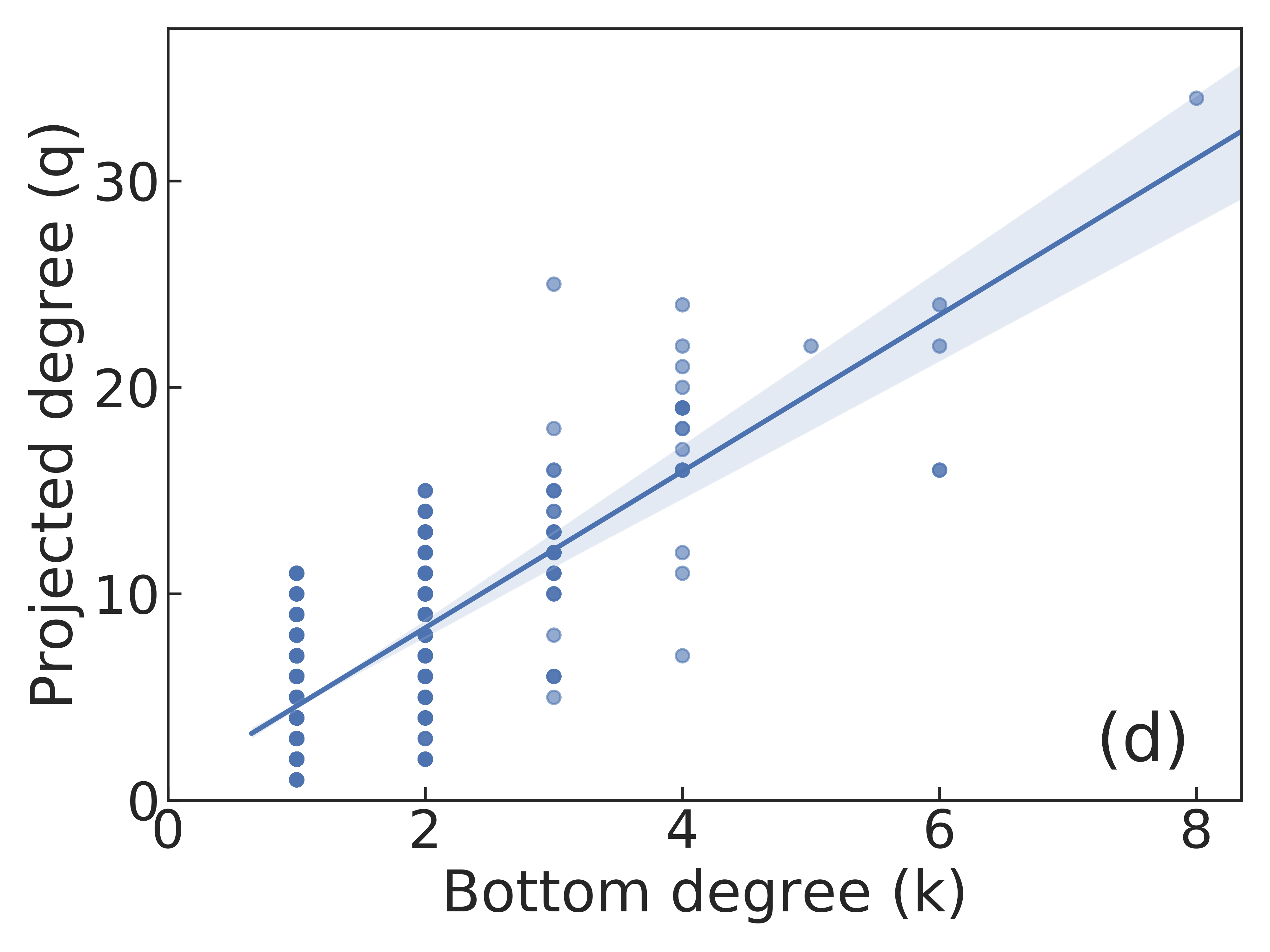}}
	\caption{Correlation between the bottom and projected degree for: (a) ArXiv Biology, (b) ArXiv Mathematics (c) PRE --- Physical Review E, and (d) Norwegian directors. Although many low degree bottom nodes have high projected degree (i.e. connected to high top degree nodes), especially in (a), we see a positive correlation between $k_{u}$ and $q^{\textrm{s}}_{u}$. That is, the higher $k_{u}$, the more likely that $u$ will have high $q^{\textrm{s}}_{u}$.}
	\label{fig:bottomvsprojected}
\end{figure*}

\begin{figure*}
	\centering
	\subfigure{\label{fig:bio_bf} \includegraphics[scale=0.19]{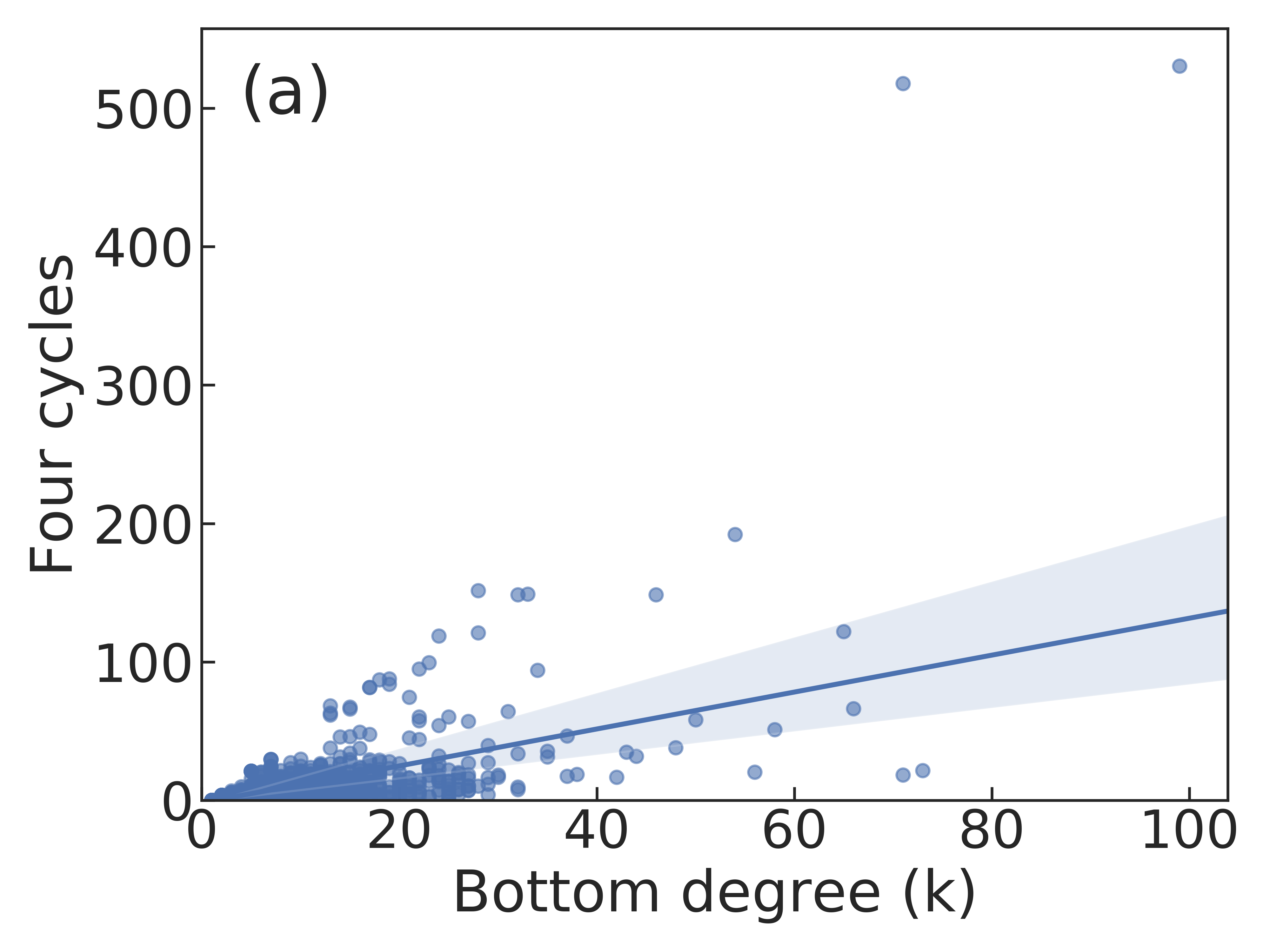}}
	\subfigure{\label{fig:maths_bf} \includegraphics[scale=0.19]{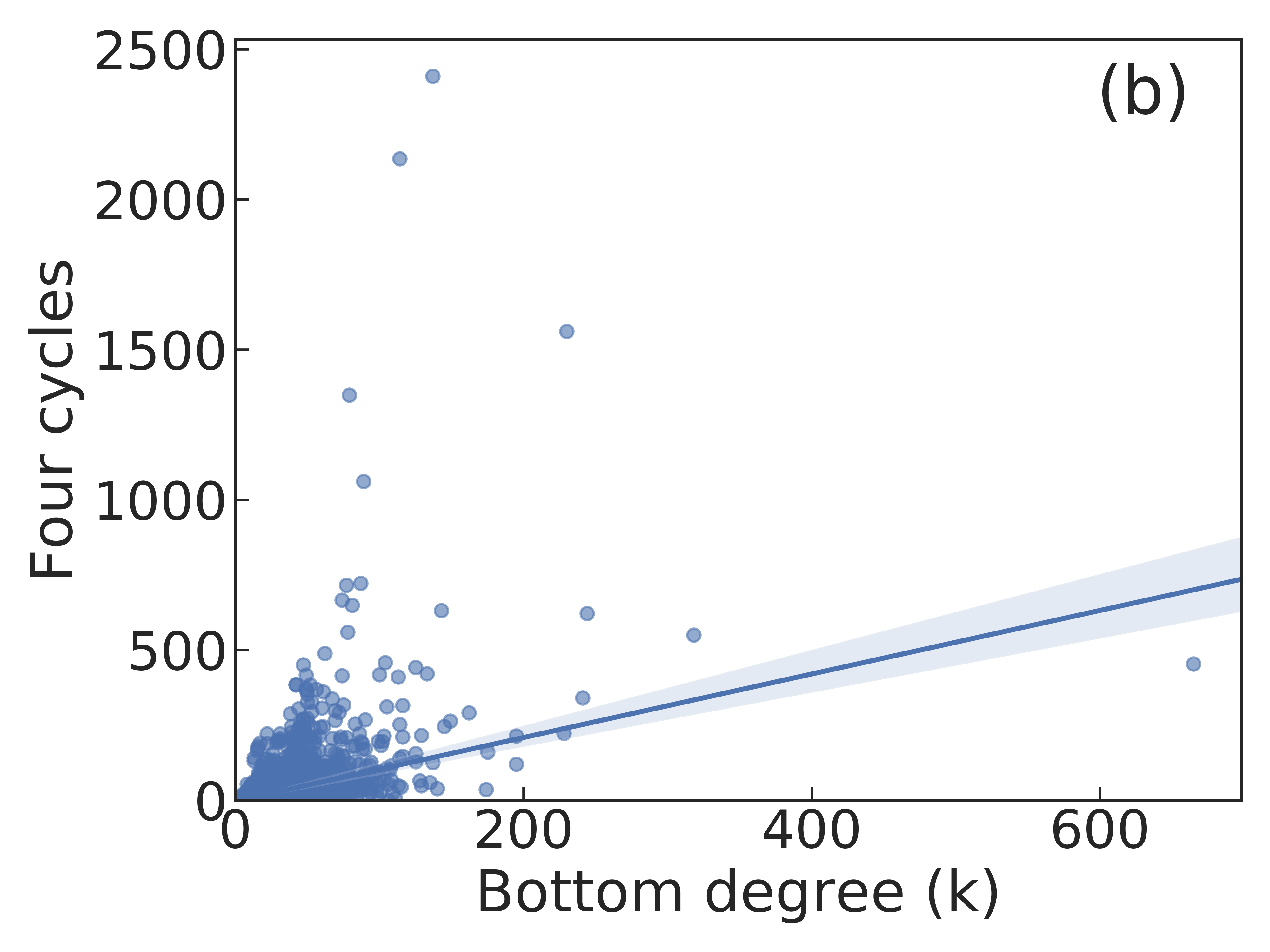}}
	\subfigure{\label{fig:pre_bf} \includegraphics[scale=0.19]{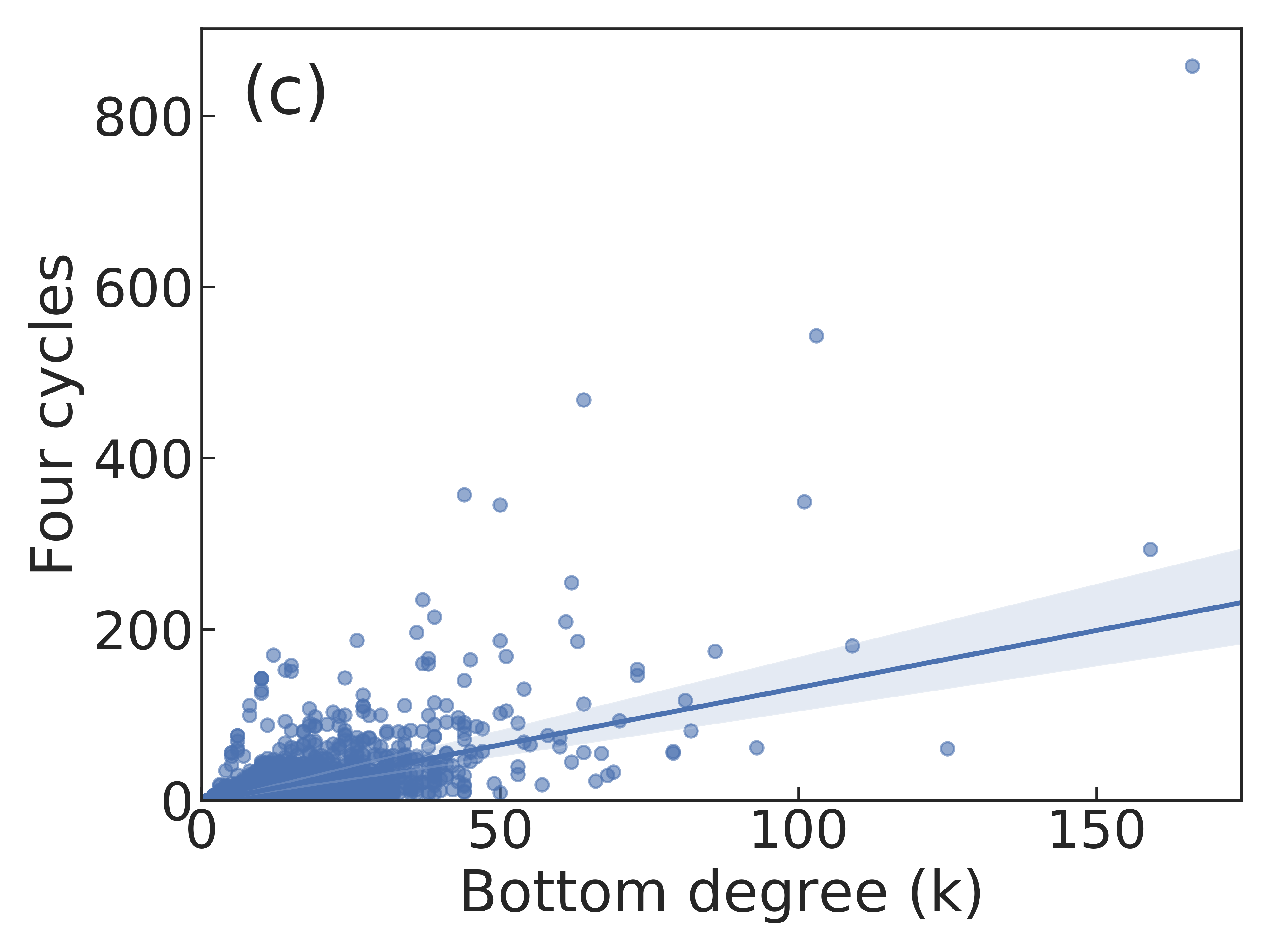}}
	\subfigure{\label{fig:norwegian_bf} \includegraphics[scale=0.19]{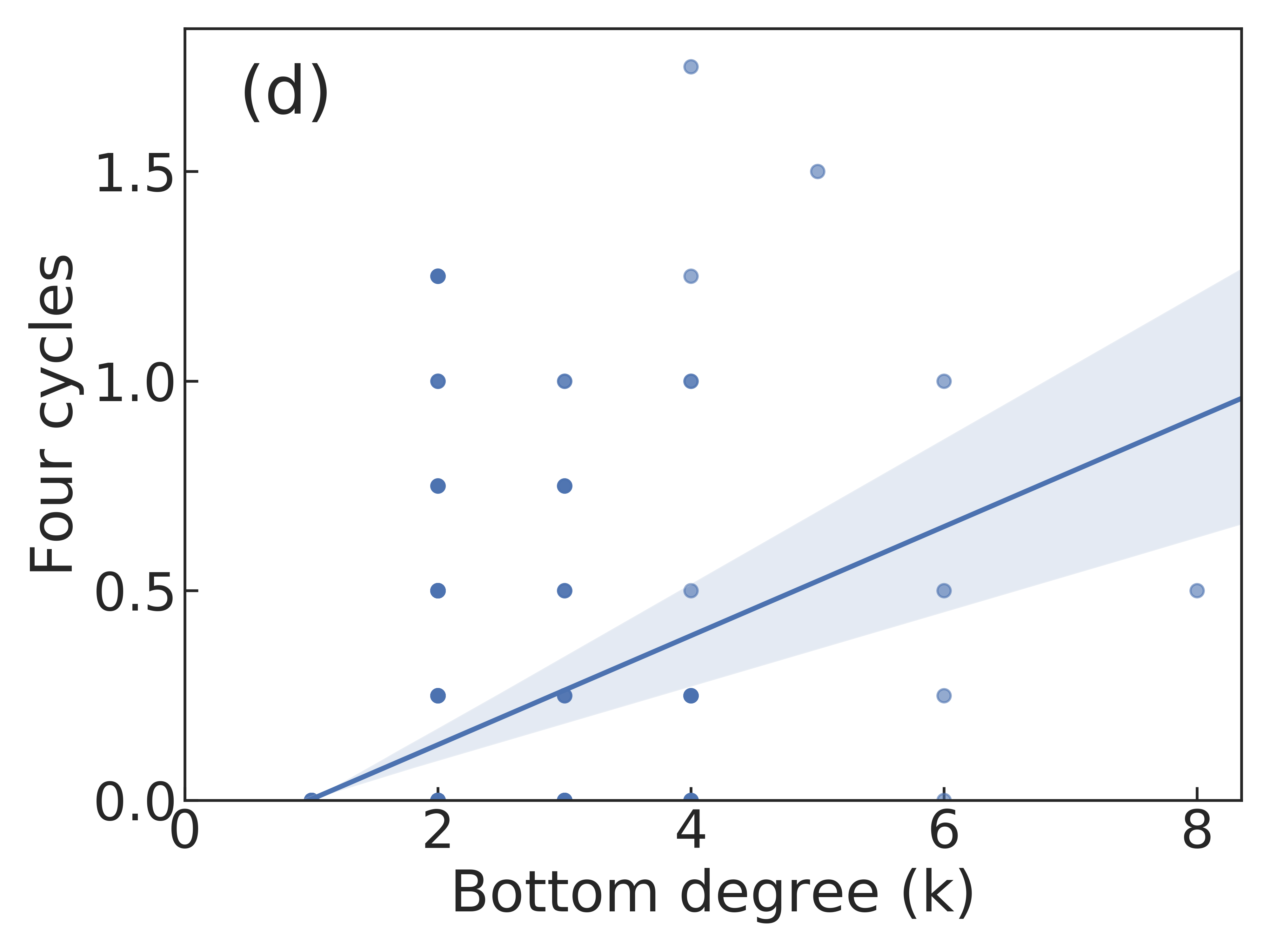}}
	\caption{Correlation between the bottom degree and the number of four-cycles in $B$ a node is part of for: (a) ArXiv Biology, (b) ArXiv Mathematics (c) PRE --- Physical Review E, and (d) Norwegian directors. The positive correlations mean that high degree bottom nodes are more likely to be part of several four-cycles (to have several common neighbors with other bottom nodes).}
	\label{fig:bottomvsfour}
\end{figure*}

From this first observation, we can say that, as top and bottom node degrees are uncorrelated, the bottom degree $k_{u}$ of node $u$ plays a major role in the resulting projected degree (and strength). That is, $u$ is more likely to have high  $q^{\textrm{s}}_{u}$ (and $q^{\textrm{m}}_{u}$) if it has high $k_{u}$ in $B$ (Fig. \ref{fig:bottomvsprojected}). However, high degree bottom nodes are also more likely to be part of more four-cycles in the bipartite network (Fig. \ref{fig:bottomvsfour}), increasing the difference between $q^{\textrm{s}}_{u}$ and $q^{\textrm{m}}_{u}$, in the projection. As consequence, the node strength distribution is more right-skewed than the degree distribution \cite{vasques2019bipartite}. We conjecture that this is why degree assortativity is higher than the node strength assortativity in the projections of bipartite networks with heavy-tailed top distributions (ArXiv Bio, ArXiv Maths, and PRE), but the same for the Norwegian network. The latter has peaked distributions (Fig. \ref{fig:norwegian}), and less four-cycles per node, with a smaller correlation slope than the others (Fig. \ref{fig:bottomvsfour}).

Continuing on degree distributions, a heavy-tailed degree distribution of top nodes is expected to generate high levels of transitivity, as aforementioned, and of degree assortativity \cite{vasques2018degree}, restricted however, by the tail of the bottom distribution. The heavier tail of the top distribution in the ArXiv Bio network for instance, induces more triangles than the top distributions of ArXiv Maths and PRE. At the same time, its more peaked bottom distribution induces less open triplets than the other two bottom distributions. As expected, ArXiv Bio has much larger transitivity $C=0.84$, than ArXiv Maths ($C=0.35$) and PRE ($C=0.51$). Yet, although the ArXiv Bio top distribution results in high relative branching ($|P_{2/1}|=24.5$), it does much more so to intermodular connectivity ($|P_{3/2}|=72.4$). The result of $|P_{3/2}| > |P_{2/1}|$ and high $C$ is large values of degree assortativity $r_{G_{\textrm{s}}}$ (Eq. \ref{eq:estrada_r}). As expected, assortativity is larger in ArXiv Bio, than in ArXiv Maths and PRE. For the Norwegian network, which displays peaked top and bottom distributions, $|P_{3/2}| = |P_{2/1}|$, thus transitivity drives $r_{G_{\textrm{s}}}$. Although both distributions are peaked, the top distribution is more shifted to the right, i.e. relatively large number of triangles compared to the number of open triplets. As consequence, transitivity is high, and the projected network is degree assortative, with $r_{G_{\textrm{s}}} = 0.17$


Let us now compare the empirical networks with the BiCM, to have a clear understanding of the role of the small cycles motifs affecting transitivity and assortativity, apart from the effects of $P_{t}(d)$ and $P_{b}(k)$. The process of randomly rewiring links of the BiCM breaks structural patterns found on empirical networks. At a first glance, one could expect that this would always reduce the number of triangles when projecting, as six-cycles are broken and the second term of Eq. \ref{eq:triangles} is smaller. However, not only six-cycles are broken in the BiCM, but also four-cycles and other patterns similar to those in Fig. \ref{fig:inhibt}. As we can see in Table \ref{tb:summary}, for every case, the number of six-cycles in the BiCM diminishes significantly, when compared to the empirical value, but the number of triangles in the projection actually grows. Nonetheless, the expansion of the number of open triplets in the projection, due to the BiCM rewiring, is much higher than any variation of the number of triangles. Thus, transitivity in BiCM projections is smaller than empirical networks for every case, as the major consequence of these small-cycles motifs (concerning transitivity) is the suppression of open triplets.

Likewise, regarding assortativity, small cycles are suppressing relative branching. To explain this, let us work out an example. Extremely heavy-tail top degree distributions, as for instance following a power-law $P_{t}(d) \propto k^{- \gamma}$, with $\gamma \approx 2.0$, will create some huge cliques in the projected network, increasing transitivity, intermodular connectivity and relative branching. If the bottom distribution is peaked, intermodular connectivity is higher, as the cliques in the projection are likely to be linked together, thus the network is assortative. However, if the bottom degree distribution is extremely heavy-tailed, as the top degree distribution, projected star-like structures (i.e. a bottom node connected to several top nodes with degree $d_{v}=2$ in $B$) are more common. Then, cliques are more distant of each other which, in turn, can result in a dissortative projection as relative branching can dominate. The presence of four-cycles helps to prevent the appearance of star-like structures and that cliques become too separated, diminishing relative branching and increasing the value of $r_{G_{\textrm{s}}}$, from what one would be expect from the observed degree distributions $P_{t}(v)$ and $P_{b}(k)$. In \cite{newman2003social}, the authors stated that their model --- based only on the degree distributions of bipartite networks (group structure) --- explains only part of the observed assortativity of the case studied. They say they ``conjecture that this is due to true sociological or psychological effects in the way in which acquaintanceships are formed.'' Based on the results presented here, we claim that four and six-cycles represent, in the network, these sociological or psychological effects. While four-cycles are repeated interactions between a pair of agents (for instance based on trust, respect, empathy, and so on), six-cycles represent triadic closure \cite{simmel2015soziologie,granovetter1973strength}.

\begin{figure}
	\centering
	\subfigure{\label{fig:pre_degree} \includegraphics[scale=0.25]{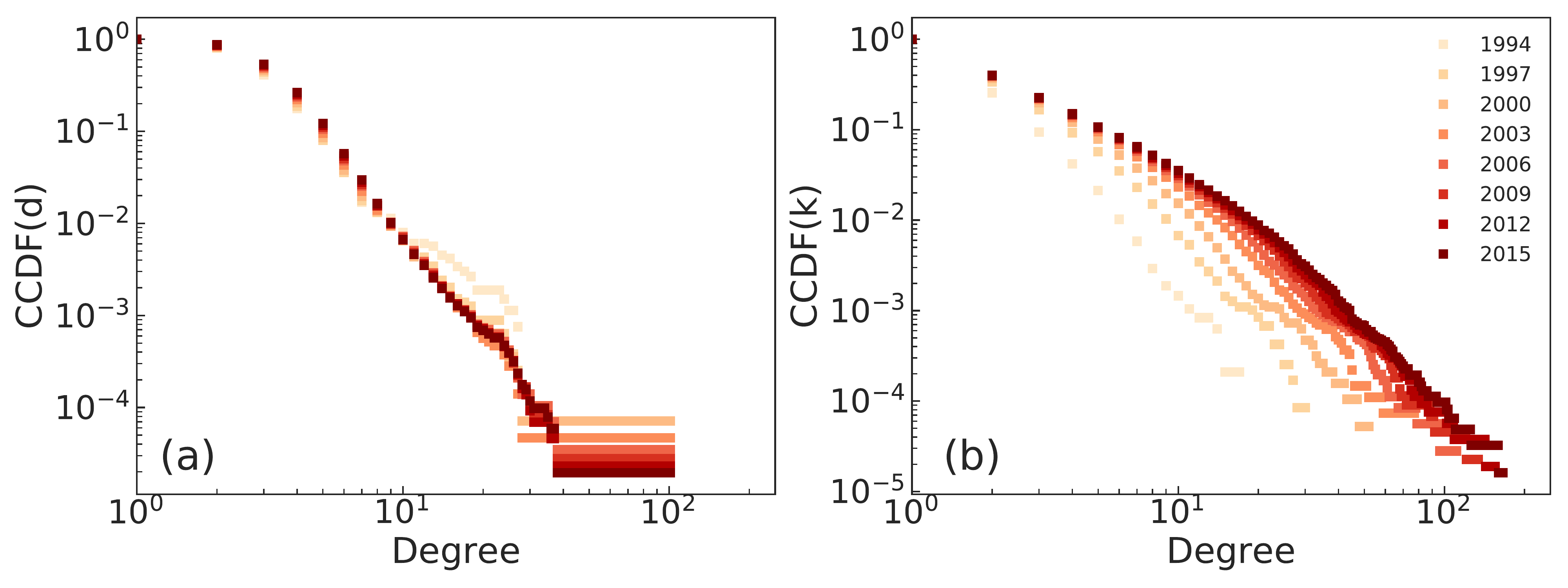}}
	\subfigure{\label{fig:pre_prop} \includegraphics[scale=0.25]{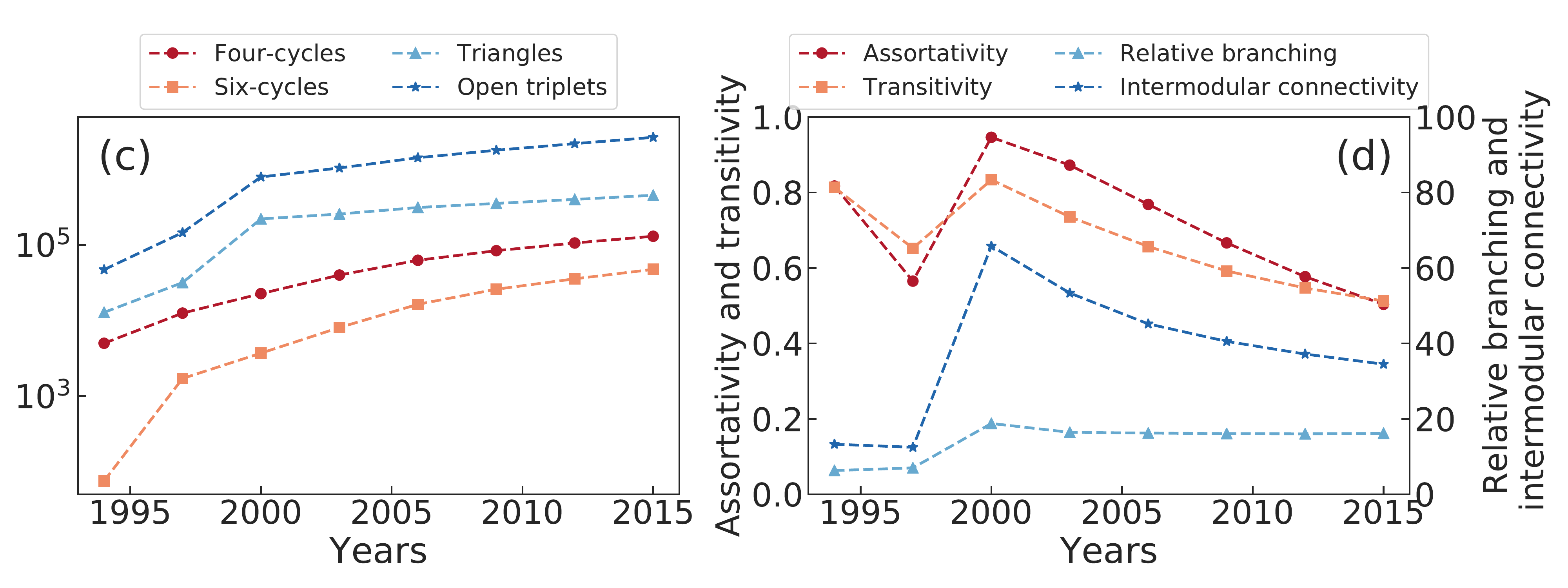}}
	\caption{Evolution of PRE network topology until 2015. (a) Top and (b) bottom degree distribution; (c) four and six-cycles, triangles, and open triplets; and (d) transitivity, assortativity, relative branching, and modular connectivity. The highest-degree top node enters the network (snapshot of year 2000), increasing transitivity and assortativity as it favors triangles and intermodular connectivity, respectively. As only the bottom degree distribution gets shifted to the right with time, open triplets and relative branching are favored, reducing transitivity and assortativity as the network grows.}
	\label{fig:pre_evolution}
\end{figure}

In order to support these findings, lets take a particular case --- the PRE network --- and investigate the evolution of the network and its topology. To this end, we created eight cumulative snapshots from 1994 to 2015, with three years intervals. We can notice that between 1997 and 2000, the highest-degree top node (the paper with the largest number of authors) enters the network, dramatically right-skewing the top degree distribution (Fig. \ref{fig:pre_evolution}(a)). The consequence for the transitivity and assortativity are visible. The frequency of triangles increases, more than the number of open triplets, increasing transitivity (Figs. \ref{fig:pre_evolution}(c) and \ref{fig:pre_evolution}(d)). Similarly, intermodular connectivity spikes, more than the relative branching, making assortativity reach an extremely high level of $0.95$. As the years go by, the tail of the top distribution does not get any heavier, while the bottom distribution keeps being shifted to the right (Fig. \ref{fig:pre_evolution}(b)). This behavior widens the gap between open triplets and triangles, and shortens the difference between intermodular connectivity and relative branching, slowly decreasing both transitivity and assortativity.    

\begin{figure}
	\centering
	\subfigure{\label{fig:aps_degree} \includegraphics[scale=0.25]{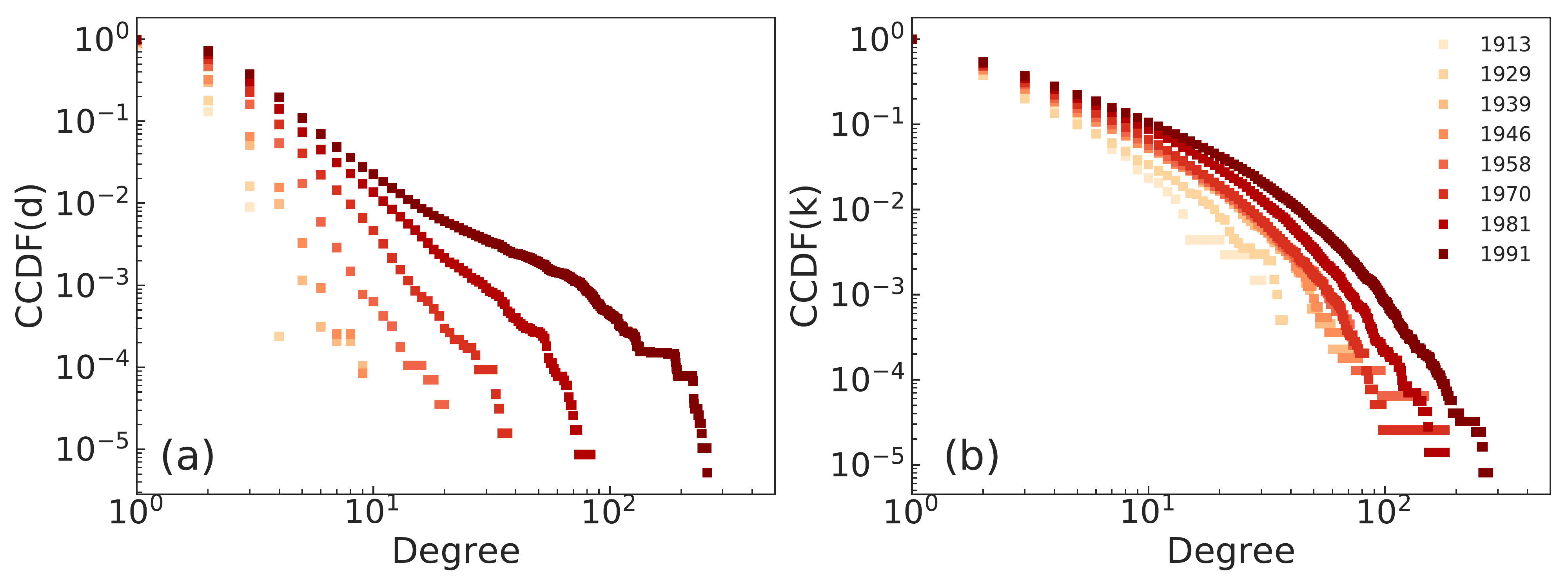}}
	\subfigure{\label{fig:aps_prop} \includegraphics[scale=0.25]{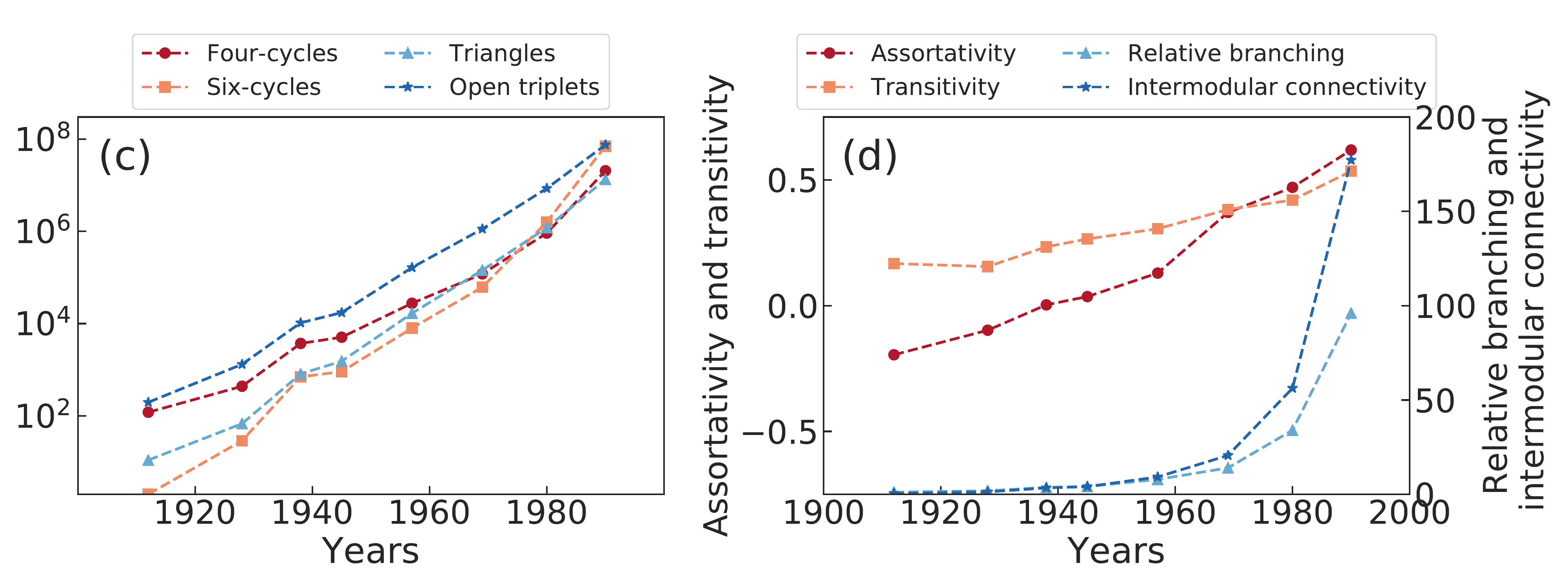}}
	\caption{Evolution of APS network topology until 1991. (a) Top and (b) bottom degree distribution; (c) four and six-cycles, triangles, and open triplets; and (d) transitivity, assortativity, relative branching, and modular connectivity. Initially the network is dissortative, due to the combination of a peaked top degree distribution, a broader bottom distribution than the top one, and the low presence of four and six-cycles. As the network evolves, the top distribution becomes broader and the frequency of small cycles grows substantially, resulting in large transitivity and degree assortativity}
	\label{fig:aps_evolution}
\end{figure}

Lastly, we would like to answer the following question: is it possible to have a one-mode degree dissortative projected network? Fisher \textit{et al.} \cite{fisher2017perceived} already presented a few examples where the assortativity of networks created with group-based methods is negative. Moreover, we have shown in \cite{vasquesfilho2019structure}, that in its ``early'' days (around 40 years), the American Physical Society co-authorship network (when Physical Review was the only journal of the society) is dissortative. Fig. \ref{fig:aps_evolution} shows the evolution of the APS network topology, from 1893 to 1991 (for the complete evolution of this network, until 2015, see \cite{vasquesfilho2019structure}). 

We can see that, until 1929, the top distribution is highly peaked and the bottom distribution has a heavier tail. The small cycles keep relative branching in relatively low levels until then, but that is not enough, as we still have $|P_{2/1}| > |P_{3/2}| + C$. Thus, the network is dissortative. As the network grows, the top distribution becomes more and more heavy-tailed, with high presence of four-cycles --- perhaps the sociological or psychological effect of researchers repeatedly writing papers with the same collaborators, and also closing open triplets, driving the network to become degree assortative.


In summary, we have shown that one-mode social networks display higher transitivity and degree assortativity than non-social networks due to the structural properties of the underlying bipartite network, but not simply because they are bipartite. We can claim that every social network has a bipartite nature, be it an affiliation, a membership, or the simplest realization of an event like accepting a friendship request on Facebook. The main driver of the topology of a projected one-mode network is the combination between the degree distributions of top and bottom nodes in the bipartite network, supported by small cycles motifs, i.e. four and six-cycles. The understanding of how these important topological features --- transitivity and degree assortativity --- emerge can be useful to improve studies and models of spreading phenomena on social networks, especially if group-based (bipartite) structures are considered.

\section*{Acknowledgments}
The authors would like to thank Chakresh Singh for helpful conversations.

\bibliographystyle{unsrt}  
\bibliography{bipsocialnetworks}

\end{document}